\documentclass[11pt]{article}

\usepackage{aaspp4}

\newcommand{\kms}{km~s$^{-1}$}
\newcommand{\vlsr}{\ensuremath{v_{\rm LSR}}}

\newcommand{\hi}{\ion{H}{1}}
\newcommand{\hii}{\ion{H}{2}}
\newcommand{\ha}{H$\alpha$}
\newcommand{\hb}{H$\beta$}
\newcommand{\sii}{\ion{S}{2}}
\newcommand{\nii}{\ion{N}{2}}
\newcommand{\oiii}{\ion{O}{3}}

\newcommand{\oi}{\ion{O}{1}}

\newcommand{\iha}{\ensuremath{I_{\mathrm{H}\alpha}}}

\newcommand{\pma}[2]{{
    \renewcommand{\arraystretch}{0.5}
    \begin{array}{ll}
      {\scriptstyle +#1} \\
      {\scriptstyle -#2}
    \end{array}}
  }
  
\begin{document}

\title{WHAM Observations of \ha, [\sii], and [\nii]
  toward the Orion and Perseus Arms: Probing the Physical Conditions
  of the Warm Ionized Medium}

\author{L. M. Haffner} \author{R. J. Reynolds} \author{S. L. Tufte}
\affil{Department of Astronomy, University of Wisconsin--Madison, 475
  North Charter Street, Madison, WI 53706\\
  Electronic Mail: {\tt haffner@astro.wisc.edu, reynolds@astro.wisc.edu,
  tufte@astro.wisc.edu}}

\begin{abstract}
  A large portion of the Galaxy ($\ell = 123\arcdeg$ to $164\arcdeg$,
  $b = -6\arcdeg$ to $-35\arcdeg$), which samples regions of the Local
  (Orion) spiral arm and the more distant Perseus arm, has been mapped
  with the Wisconsin H-Alpha Mapper (WHAM) in the [\sii] $\lambda
  6716$ and [\nii] $\lambda 6583$ lines. By comparing these data with
  the maps from the WHAM \ha\ Sky Survey, we begin an investigation of
  the global physical properties of the Warm Ionized Medium (WIM) in
  the Galaxy.  Several trends noticed in emission-line investigations
  of diffuse gas in other galaxies are confirmed in the Milky Way and
  extended to much fainter emission. We find that the [\sii]/\ha\ and
  [\nii]/\ha\ ratios increase as absolute \ha\ intensities decrease.
  For the more distant Perseus arm emission, the increase in these
  ratios is a strong function of Galactic latitude, $b$, and thus, of
  height, $z$, above the Galactic plane, while the [\sii]/[\nii] ratio
  is relatively independent of \ha\ intensity. Scatter in this ratio
  appears to be physically significant, and maps of [\sii]/[\nii]
  suggest regions with similar ratios are spatially correlated. The
  Perseus arm [\sii]/[\nii] ratio is systematically lower than Local
  emission by 10\%--20\%. With [\sii]/[\nii] fairly constant over a
  large range of \ha\ intensities, the increase of [\sii]/\ha\ and
  [\nii]/\ha\ with $|z|$ seems to reflect an increase in temperature.
  Such an interpretation allows us to estimate the temperature and
  ionization conditions in our large sample of observations. We find
  that WIM temperatures range from 6,000 K to 9,000 K with temperature
  increasing from bright to faint \ha\ emission (low to high
  [\sii]/\ha\ and [\nii]/\ha) respectively. Changes in [\sii]/[\nii]
  appear to reflect changes in the local ionization conditions
  (\emph{e.g.}\ the S$^+$/S$^{++}$ ratio). We also measure the
  electron scale height in the Perseus arm to be $1.0\pm0.1$ kpc,
  confirming earlier, less accurate determinations.
\end{abstract}


\section{Introduction}
\label{sec:intro}

Although the Wisconsin H-Alpha Mapper (WHAM) Sky Survey will present
the first view of the distribution and kinematics of the Warm Ionized
Medium (WIM), we still know little about the physical conditions in
this gas or how they are produced.  Several studies have measured
constraints on important parameters toward specific regions of the
WIM, but none has yet provided a large sample of measurements to help
test how environmental conditions (\emph{e.g.}\ distance from \hii\ 
regions, distance from the Galactic plane, etc.) might affect these
physical conditions.
 
Reynolds (1985a\markcite{rjr85a}) derived the first estimate of the
temperature of the WIM by combining measured line widths of [\sii] and
\ha\ emission toward the same regions. He was able to place limits
between 5,000 K and 20,000 K, with a mean of about 8,000 K, on the
temperature of a sample of 21 WIM observations. Ionization conditions
have been probed even less, with most of our current information
coming from observations in the plane of the Galaxy. Reynolds
(1985b\markcite{rjr85b}) showed that [\oiii] $\lambda 5007$ emission
is very faint in the WIM, placing limits on the amount of O$^{++}$
(needing $h\nu > 35.1$ eV) in the WIM.  More conclusively, Reynolds \&
Tufte (1995\markcite{rt95}) and Tufte (1997\markcite{t97}) have shown
via observations of the He I $\lambda 5876$ recombination line that
the fraction of He$^{+}$ (needing $h\nu > 24.6$ eV) in this gas is
also quite low in comparison with traditional O star \hii\ regions.
The recent detection of [\oi] $\lambda 6300$ from three WIM directions
(Reynolds, et al.\ 1998\markcite{rhth98}) places further limits on the
ionizing spectrum. The low [\oi]/\ha\ ratios seen in all three
directions suggests that the H is mostly H$^{+}$ in the \ha\ emitting
gas.

The advent of WHAM provides a unique opportunity to trace some of
these diagnostics over a larger region of the sky. This paper presents
the first velocity-resolved study of [\sii] and [\nii] from the
diffuse gas of our Galaxy. A region of the sky that includes the
Perseus arm was chosen for examination since the \ha\ emission from
the WIM in this arm is well separated in velocity from local WIM
emission and can be traced to high Galactic latitude, allowing an
exploration of the emission with distance, $z$, from the Galactic
plane. Such observations can be directly compared to studies of
edge-on galaxies, most notably those by Rand (1997\markcite{rand97},
1998\markcite{rand98}) of NGC 891. The primary trends found in his
work are borne out here and extended to even fainter emission
measures. Specifically, [\sii]/\ha\ and [\nii]/\ha\ ratios increase
with $|z|$, while [\sii]/[\nii] ratios remain nearly constant.

A brief description of the instrument and observations are presented
in \S\ref{sec:obs}. The results from the data, including a measurement
of the electron scale height in the Perseus arm and an investigation
of the trends in line ratios, are presented in \S\ref{sec:results}. In
\S\ref{sec:discussion}, we present our interpretation that these line
ratios provide direct measurements of the temperature and ionization
state of the WIM. Finally, our conclusions are summarized in
\S\ref{sec:summary}.

\section{Observations}
\label{sec:obs}

\subsection{The WHAM Instrument}
\label{sec:wham}

Due to the extended nature of the WIM, large-aperture Fabry-Perot
detection techniques have proven to be quite successful in studying
the WIM (Reynolds et al.\ 1990\markcite{rrsh90}). For WHAM, a 6-inch,
dual-etalon Fabry-Perot spectrometer delivers the high-efficiency,
high-resolution observations (Tufte 1997\markcite{t97}; Reynolds et
al.\ 1998\markcite{rthjp98}). The dual-etalon design provides superior
order rejection over single-etalon systems to achieve a much larger
free spectral range while maintaining high spectral resolution (12
\kms). An all-sky siderostat housing a 0.6-meter lens guides light to
the etalon chambers in an environmentally controlled trailer. The
center wavelength of the 200 \kms\ spectral window is selected by
changing the pressure of a high index of refraction gas (SF$_6$) in
sealed etalon chambers. A narrow-band ($\sim 20$ \AA\ FWHM)
interference filter provides additional spectral isolation near the
desired wavelength. The resulting spectrum, in the form of a
Fabry-Perot ring pattern, is imaged onto a cryogenically cooled Tek
1024$\times$1024 CCD.

The spectrum covers a 200 \kms\ interval at 12 \kms\ spectral
resolution. Coatings on the etalons and optics allow the placement of
the 200 \kms\ wide spectral window anywhere between 4800 and 7300 \AA.
In this ``spectral mode,'' the image contains only the average
spectrum within the one-degree diameter circular beam, avoiding
confusion between spectral features and angular structures of the
source (or stars) within the beam. WHAM also has a set of optics that
can be placed in the post-etalon beam to provide an extremely
narrow-band (adjustable between 12-200 \kms), 1\arcmin\ angular
resolution \emph{image} of the sky within a one-degree field of view.
Only the spectral mode (with one-degree angular resolution) is used
for the observations presented here. Further details on the WHAM
spectrometer and its calibration can be found in Tufte
(1997\markcite{t97}).

WHAM is currently located at Kitt Peak National Observatory. Using the
WIYN messaging system designed by Percival (1994\markcite{p94},
1995\markcite{p95}) and custom hardware designed by the {\em Space
  Astronomy Lab} at Wisconsin, WHAM is a completely remotely operated
telescope system. All functions of the siderostat and spectrometer are
routinely operated from our Wisconsin offices.  Details on the
hardware and software used to operate WHAM and to automate the survey
can be found in Haffner (1999\markcite{h99}).

\subsection{The Data}
\label{sec:data}

The \ha\ observations of the 1100 deg$^2$ region $\ell = 123\arcdeg$
to $164\arcdeg$, $b = -6\arcdeg$ to $-35\arcdeg$ that is the focus of
this study were obtained between 1997 July and 1997 October as a part
of the WHAM \ha\ Sky Survey (Haffner 1999\markcite{h99}). Additional
observations at the same survey grid points but with the 200 \kms\ 
spectral window tuned near the [\sii] $\lambda 6716$ and [\nii]
$\lambda 6583$ lines were obtained in 1997 October, 1997 November, and
1998 June. The region presented here is sampled by 1360 \ha, 1360
[\sii], and 1360 [\nii] spectra, each representing the integrated
emission within WHAM's one-degree beam. The integration time for each
spectrum was 30 s for \ha\ and 60 s for the other two emission lines.

The [\sii] and [\nii] observations were converted into spectra using 
the same steps as those for \ha\ presented in Haffner 
(1999\markcite{h99}).  Separate white-light flat fields were created 
for each of these additional filters and used in place of the \ha\ 
flat field.  There are no strong terrestrial components in [\sii] or 
[\nii] comparable to the geocoronal emission in the \ha\ spectra.  For 
the [\sii] observations, the weak terrestrial (Tucson lights) Ne I 
line at 6717.04 \AA\ was used for velocity calibration.  For the 
[\nii] observations, spectra of \ha\ and [\nii] from bright emission 
nebulae were compared and the velocity frame of [\nii] was adjusted so 
that the mean velocity of the nebular [\nii] emission matched that of 
the \ha.  By using the terrestrial lines in the \ha\ spectra 
(geocoronal \ha) and [\sii] spectra (Ne I), the velocity scale in each 
spectrum is calibrated to better than 1 \kms.  Intensities were 
calibrated by applying a correction factor to the [\sii] and [\nii] 
observations (0.94 and 1.15, respectively) based on the transmission 
differences between these filters and the \ha\ filter.

As in the \ha\ observations, the [\sii] and [\nii] spectra are 
contaminated by weak atmospheric lines (Haffner, Reynolds, \& Tufte 
1998\markcite{hrt98}; Haffner 1999\markcite{h99}).  The brightest of 
these in our [\sii] spectra is the Ne I line located $+26.8$ \kms\ 
from the geocentric zero of [\sii] $\lambda6716$.  As noted above, it 
proves to be a useful wavelength calibrator.  To accurately 
characterize the parameters of these faint lines in the [\sii] and 
[\nii] spectra, we followed the same strategy outlined in Haffner 
(1999\markcite{h99}) for \ha.  We first created an average spectrum 
for each survey ``block'' ($\approx49$ pointings taken sequentially).  
Using this high signal-to-noise spectra, a single Gaussian was fitted 
to each of the atmospheric lines and then subtracted from the 
individual pointings.  Such corrections account for hourly variations 
in these faint lines (typically $\sim$0.02 R).  A list of the 
atmospheric lines in our [\sii] and [\nii] observations is given in 
Table~\ref{tab:atlines}.  The potential atmospheric line at $-40$ 
\kms\ in [\nii] is coincident with interstellar emission from the 
Perseus arm in all spectra presented here.  The implications of such a 
line are discussed in \S\ref{sec:discussion}.

\section{Results}
\label{sec:results}

\subsection{Overview}
\label{sec:overview}
Figure~\ref{fig:spectra} shows samples of \ha, \nii, and \sii\ spectra 
from the region under study.  The Local (centered near 0 \kms) and 
Perseus arm (centered near $-40$ to $-60$ \kms) emission are shown as 
well-separated components in the right-hand panels, where $\ell < 
150\arcdeg$, while Local gas dominates in the left panels ($\ell > 
150\arcdeg$).  Figures~\ref{fig:loc-lr}a and \ref{fig:per-lr}a display 
\ha\ intensity maps integrated over $\vlsr = -10$ to $+10$ \kms\ 
(Figure~\ref{fig:loc-lr}) and $\vlsr = -30$ to $-50$ \kms\ 
(Figure~\ref{fig:per-lr}), sampling the Local and Perseus arms, 
respectively.  The boxes overplotted in them denote the $10\arcdeg 
\times 12\arcdeg$ region previously mapped in \ha\ by Reynolds 
(1980\markcite{rjr80}) and compared to \hi\ observations in detail by 
Reynolds, et al.\ (1995\markcite{rtkmh95}).  The [\sii] and [\nii] 
maps look qualitatively similar to \ha\ and are not reproduced here.  
The line ratios images [\nii]/\ha\ (Figures~\ref{fig:loc-lr}b and 
\ref{fig:per-lr}b), [\sii]/\ha\ (Figures~\ref{fig:loc-lr}c and 
\ref{fig:per-lr}c), and [\sii]/[\nii] (Figures~\ref{fig:loc-lr}d and 
\ref{fig:per-lr}d) are discussed below in \S\ref{sec:lr}.

Two large \ha\ emitting regions dominate the emission at
lower latitudes, above $b = -20\arcdeg$. The brightest emission arises
from NGC 1499 (California Nebula) excited by the O7.5I star $\xi$ Per
located $540\pma{330}{150}$ pc away (\emph{Hipparcos}; Perryman
1997\markcite{hipparcos}) at $\ell = 160\fdg37$, $b =
-13\fdg11$. Emission in the nebula itself, a $5\arcdeg
\times 2\arcdeg$ crescent centered near $\ell = 160\arcdeg$, $b =
-13\arcdeg$, peaks at over 250 R (1 R $= 10^6/4\pi$ ph cm$^{-2}$
s$^{-1}$ sr$^{-1} = 2.4\times10^{-7}$ erg cm$^{-2}$ s$^{-1}$ sr$^{-1}$
at \ha) and is centered around $\vlsr = -5$ \kms, while faint,
extended emission above 5 R centered near $\vlsr = +5$ \kms\ continues
5--7$\arcdeg$ away from the nebula. This extended \hii\ region has
been studied in some detail by Reynolds (1988\markcite{rjr88}), who
first hypothesized that the bar of emission ($I_{\mathrm{H}\alpha}
\approx 8$ R) running from $\ell = 150\arcdeg$, $b = -20\arcdeg$ to
$\ell = 142\arcdeg$, $b = -10\arcdeg$ centered between $\vlsr = -5$ to
$+5$ \kms\ may also be ionized by the star.

The second large emission region, centered near $\ell = 131\arcdeg$,
$b = -11\arcdeg$, is a superposition of separate but equally bright
emission regions in the Local and Perseus arms (see
Figure~\ref{fig:spectra}). Emission from the Local arm is near $\vlsr
= 0$ \kms\ and has a more circular appearance reminiscent of a
classical Str\"{o}mgren sphere (see Figure~\ref{fig:loc-lr}a). This
faint \hii\ region has a relatively constant brightness of 5--6 R over
most of its $13\arcdeg$ diameter area. The most promising candidate
for an ionizing source for this region is $\phi$ Per, a B0.5e + sdO
binary system located at $\ell = 131\fdg3$, $b = -11\fdg3$ and
$220\pma{43}{32}$ pc away (\emph{Hipparcos}; Perryman
1997\markcite{hipparcos}; Gies et al.\ 1998\markcite{gea98}). At that
distance, the region is 50 pc in diameter. Some details of this region
can be found in Haffner (1999\markcite{h99}). A future study (Haffner,
Reynolds, \& Tufte 1999\markcite{hrt99}) examining the properties of
B-star \hii\ regions in the WHAM \ha\ Sky Survey will expand on this
work.  Emission from the Perseus arm in this direction at $-60 < \vlsr
< -40$ is much more amorphous in shape (Figure~\ref{fig:per-lr}a).

At higher latitudes and at velocities farther from the LSR, the \ha\ 
emission becomes quite filamentary. Some of the more notable
structures include a faint bar in Figure~\ref{fig:per-lr}a running
from $\ell = 150\arcdeg$, $b = -33\arcdeg$ to $\ell = 160\arcdeg$, $b
= -25\arcdeg$; several, nearly vertical filaments near $\ell =
130\arcdeg$, $b = -20\arcdeg$; and the shell-like region and
associated dissecting filament (more visible at higher velocities; see
Haffner 1999\markcite{h99}) centered on the box in
Figure~\ref{fig:per-lr}a. The last of these was previously studied in
detail by Ogden \& Reynolds (1985\markcite{or85}) and Reynolds et al.\ 
(1995\markcite{rtkmh95}).

\subsection{Scale Height of the WIM}
\label{sec:hvsb}

Since the Perseus arm gas is well separated in velocity from the Local
emission, we can explore the decrease of this emission as a function
of the height above the Galactic plane. To extract information about
the scale height from the intensity distribution, we first assume that
the gas density is primarily a function of $z$ and has the form
\begin{equation}
  \label{eq:gasexp}
  n_e(z) = n_e^0\;e^{-|z|/H}\;\mathrm{cm}^{-3},
\end{equation}
where $n_e^0$ is the mid-plane density and $H$ is the scale-hight of
the WIM. The \ha\ intensity is related to the gas density by the
equation
\begin{equation}
  \label{eq:itoem}
  2.75\;T_4^{0.9}\,\iha = \int \phi\;n_e^2\;dl,
\end{equation}
where $\phi$ is the filling fraction of the emitting gas, $T_4$ is the
temperature of the gas in $10^4$ K, and $\iha$ is in R. As long as the
temperature, filling factor, and pathlength through the emitting gas,
$\int dl = L$, are not functions of $z$ we then have
\begin{equation}
  \label{eq:ivsz}
  \iha = \frac{\phi\,(n_e^0)^2\,L}{2.75\,T_4^{0.9}}\;e^{-2|z|/H} =
  \iha^0\;e^{-2|z|/H},
\end{equation}
where $\iha^0$ is the intensity at the mid-plane. Finally, 
substituting $z = D\,\tan |b|$, where $D$ is the distance to the arm and
$b$ is the Galactic latitude of an observation, and taking the natural
log, we arrive at
\begin{equation}
  \label{eq:ivstanb}
  \ln \iha = \ln \iha^0 - \frac{2D}{H} \tan |b|.
\end{equation}

Figure~\ref{fig:havstanb} plots this relationship for a selected 
region of the map.  Each point represents the median emission between 
$\ell=125\arcdeg$ and $152\arcdeg$ for each Galactic latitude.  The 
median of the emission in each latitude slice is used instead of an 
average to reduce the effect of local emission enhancements 
(\emph{e.g.}\ faint \hii\ regions).  The limited longitude range is 
chosen to avoid the gradual blending of the Perseus arm component with 
the Local gas as the longitude approaches $180\arcdeg$.  Three radial 
velocity intervals are presented in the plot, reflecting total ($\vlsr 
= -100$ to $+100$ \kms), Local ($\vlsr = -25$ to $+100$ \kms), and 
Perseus arm ($\vlsr = -100$ to $-25$ \kms) emission.  The integration 
ranges are expanded from the limited ranges in 
Figures~\ref{fig:loc-lr} and \ref{fig:per-lr} to include the full emission 
profile in our calculations.  The error bars for the Perseus arm are 
an estimate of the physical scatter along a latitude slice calculated 
from the average deviation about the median of emission within that 
slice.

Also displayed in Figure~\ref{fig:havstanb} is a linear fit to
Perseus arm points that are above $b = -15\arcdeg$. Closer to the
plane, \hii\ regions and extinction play a significant role and these
points are excluded from the fit. The resulting fit parameters are
$\ln \iha^0 = 1.73\pm0.19$ and $2D/H = 4.91\pm0.39$, leading to a
mid-plane intensity for the WIM of $5.7\pm0.2$ R and a scale height of
$1.0\pm0.1$ kpc if the Perseus arm is at $D = 2.5$ kpc (see Reynolds et
al.\ 1995\markcite{rtkmh95} and references therein). The dotted line
in Figure~\ref{fig:havstanb} presents a manual fit to the lower
envelope of the Perseus arm emission. The resulting parameters are
essentially the same: $H = 1.0$ kpc and $\iha^0 = 4.2$ R. The 1.0 kpc
scale height appears to be valid for $|z|$ from 750 pc to at least
1700 pc, the boundary of this studied region. 

As mentioned above, this estimate assumes that the filling fraction 
and temperature do not depend on $|z|$.  If in fact $\phi$ changes as 
a function of height above the plane, it affects our estimate of the 
electron scale height.  As shown below in \S\ref{sec:discussion}, 
these new observations suggest that the temperature in the WIM rises 
with distance from the Galactic plane.  Such an effect would increase 
our estimate of the scale height.  Finally, we also note here that the 
full role of interstellar extinction has not been taken into account 
within our restricted fitting region.  Future observations of H$\beta$ 
will allow a straightforward correction of the H$\alpha$ data for 
this attenuation.  A simplistic estimate of the magnitude of this 
correction within our fitted region suggests a decrease in the 
inferred scale height of $\sim10\%$, somewhat offsetting the expected 
effect of $T(|z|)$.

\subsection{Line Ratios}
\label{sec:lr}

Figures~\ref{fig:loc-lr} and \ref{fig:per-lr} also present
pseudo-color images of (b) [\nii]/\ha, (c) [\sii]/\ha, and (d)
[\sii]/[\nii] line ratios integrated over $\vlsr = -10$ to $+10$ \kms\ 
(Figure~\ref{fig:loc-lr}) and $\vlsr = -30$ to $-50$ \kms\ 
(Figure~\ref{fig:per-lr}), sampling the Local and Perseus arms,
respectively. The [\nii]/\ha\ and [\sii]/\ha\ maps in both arms
generally anti-correlate with \ha\ emission except for the $\phi$ Per
\hii\ region centered near $\ell = 131\arcdeg$, $b = -11\arcdeg$
mentioned in \S\ref{sec:overview} above. [\sii]/[\nii] ratios are
generally lower in the the two brighter \hii\ regions in the Local
arm, but otherwise show little additional correspondence to \ha\ maps.

In Figures~\ref{fig:siivsha} and \ref{fig:niivsha}, [\sii]/\ha\ and
[\nii]/\ha\ are plotted against \ha\ intensity. The radial velocity
ranges for each panel of the figures are restricted to separate the
Local and Perseus arm emission (see Figure~\ref{fig:spectra}). As
above, the longitudinal extent of the Perseus observations is limited
to $\ell < 152\arcdeg$. We denote the pointings within the newly
discovered $\phi$ Per \hii\ region (see \S\ref{sec:overview}) with a
square symbol. Note that these points appear to form the upper
envelope of the [\sii]/\ha\ ratios for their \ha\ intensities
(Figure~\ref{fig:siivsha}) and actually form a distinctly different
horizontal branch in the [\nii]/\ha\ ratios
(Figure~\ref{fig:niivsha}). In general, both figures show a marked
increase in these ratios with decreasing \ha\ intensity. For
comparison to a more traditional \hii\ region in this mapped area, the
horizontal, solid line at the right edge of both figures shows the
average ratio from the six pointings with $\iha > 50$ R near NGC 1499:
[\sii]/\ha\ = 0.14 and [\nii]/\ha\ = 0.39.  Ratios for a brighter
\hii\ region, our calibration source NGC 7000 ($\iha = 800$ R), are
[\sii]/\ha\ = 0.06 and [\nii]/\ha\ = 0.20 (see also Reynolds
1985a\markcite{rjr85a}).

Figure~\ref{fig:siivsnii} plots these two line ratios against one
another. In both the Local and Perseus arms, the ratios appear to be
strongly correlated. In this figure, the $\phi$ Per (B-star) and $\xi$
Per (O-star) \hii\ regions are highlighted with square and triangle
symbols, respectively. Both appear to occupy specific loci that are
different not only from one another, but from the general WIM
background. The Perseus arm panel is again restricted to $\ell <
152\arcdeg$.

As discussed in \S\ref{sec:hvsb}, the general trend in \ha\ intensity
is that it decreases as the observed gas lies farther from the
Galactic plane. Figure~\ref{fig:lrvb} shows that the relationship of
the [\sii]/\ha\ and [\nii]\ha\ ratios versus \ha\ intensity described
above translate directly into an increase of these ratios with
increasing $|b|$ (or $|z|$) in the Perseus arm. The figure shows the
three line ratios [\sii]/\ha, [\nii]/\ha, and [\sii]/[\nii] as a
function of $\tan |b|$. Both panels are restricted to the longitude
range $\ell = 125\arcdeg$--$152\arcdeg$ for consistency. Note also
that the [\sii]/[\nii] ratio is relatively constant, with perhaps a
slight increase in the Local arm with increasing $|b|$.

\section{Discussion}
\label{sec:discussion}

The tight correlation of [\sii] and [\nii] emission in
Figure~\ref{fig:siivsnii} results in a fairly constant value of
[\sii]/[\nii] over a wide variety of \ha\ intensities and latitudes in
both arms.  This result has been seen in numerous studies of the
extragalactic WIM analogue, commonly referred to as the diffuse
ionized gas (DIG) (Otte \& Dettmar 1999\markcite{od99}; Rand
1998\markcite{rand98}; Wang, Heckman, \& Lehnert 1997\markcite{whl97};
Greenawalt, Walterbos, \& Braun 1997\markcite{gwb97}). As Rand covers
in detail, current photoionization models in which variations in
[\sii]/\ha\ and [\nii]/\ha\ arise primarily from variations in the
ionization parameter (\emph{e.g.}\ Domg\"orgen \& Mathis
1994\markcite{dm94}; Sokolowski 1994\markcite{s94}; see also
Bland-Hawthorn, Freeman, \& Quinn 1997\markcite{bfq97}) have difficulty
keeping this ratio constant as [\sii]/\ha\ and [\nii]/\ha\ rise.
Abundance gradients and \hii\ region contamination can be a larger
problem in interpreting such ratios in extragalactic studies with the
large physical sizes represented by a typical data point. Our new
observations within the Galaxy provide smaller scales and adequate
velocity resolution to separate radial components of gas, avoiding
some of these problems.

We suggest that the tight correlation of [\sii]/[\nii] seen in our
data arises simply from the dominance of S$^{+}$ and N$^{+}$
ionization states in this gas together with the similar excitation
potential of these two emission lines. With increasing electron
temperature, [\sii]/\ha\ and [\nii]/\ha\ ratios rise together with no
change in the [\sii]/[\nii] ratio. Using this premise, namely, that
variations in these ratios are due primarily to variations in electron
temperature, we attempt to derive estimates of the electron
temperature and ionization state of S within our sample. We start with
the basic equation for the intensity of emission lines from
collisionally excited ions (Osterbrock 1989\markcite{o89}),
\begin{equation}
  \label{eq:collisional}
  I_\nu\ \mathrm{(ph\ s^{-1}\ cm^{-2}\ ster^{-1})} = \frac{f_\nu}{4\pi}\,
  \int n_i\,n_e\;\frac{8.63 \times
    10^{-6}}{T^{0.5}}\,\frac{\Omega(i,j)}{\omega_i}\,e^{-\frac{E_{ij}}
    {k T}}\,dl,
\end{equation}
where $\Omega(i,j)$ is the collision strength of the transition, 
$\omega_i$ is the statistical weight of the ground level, $E_{ij}$ is 
is the energy of the upper level of the transition above the lower, 
and $f_\nu$ is the fraction of downward transitions that produce the 
emission line in question.  We adopt the following parameterized 
representations of the collision strengths supplied by Aller 
(1984\markcite{a84}), valid from 5000 K to 20,000 K:
\begin{eqnarray}
  \label{eq:css}
  \mathrm{S^+\!:}\ \Omega(^4S_{3/2},^2D_{5/2}) & = & 4.19 \times T_4^{-0.093}\\
  \label{eq:csn}
  \mathrm{N^+\!:}\ \Omega(^3P,^1D) & = & 2.68 \times T_4^{0.026}
\end{eqnarray}
where $T_4$ is in units of $10^4$ K. A term to account for extinction
has not been included in Equation~\ref{eq:collisional}, since the
three lines we are comparing are close in wavelength.

To arrive at any estimates, we must be able to compute the path
averaged integrals in these equations. Since there is little knowledge
yet of the details of the spatial distribution of the physical
parameters in the WIM, we take the simplest approach and assume that
physical values do not vary along a path. For the ratios we calculate
below, we need to make this assumption for the temperature,
ionization, and abundance of the gas. Our estimates should then be
viewed as representative values, rather than accurately reflecting
local conditions from where the emission arises. Since we are able to
cleanly separate Local and Perseus arm gas, large-scale Galactic
variations can be probed along lines of sight, even with this
assumption. This simplification also implies that the line emission
from the three ions originates from the same location. This is
probably a valid assumption in the WIM, unless local processes on
small scales (\emph{e.g.}\ shocks) dominate the ionization structure.
In the brighter, discrete \hii\ regions, however, this assumption most
likely breaks down since the ionization structure of the nebula could
be changing significantly within our beam.

With this simplification, we arrive at these two equations for the
intensity of [\sii] and [\nii] emission,
\begin{equation}
  \label{eq:sii}
  I_{6716}\ \mathrm{(R)} = 2.79 \times 10^{5}\;\left(\frac{H^+}{H}\right)^{-1}
  \left(\frac{S^{\rule{0ex}{1ex}}}{H}\right)
  \left(\frac{S^+}{S}\right)\;T_4^{-0.593}\;e^{-2.14/T_4}\;EM,
\end{equation}
and
\begin{equation}
  \label{eq:nii}
  I_{6583}\ \mathrm{(R)} = 5.95 \times 10^{4}\;\left(\frac{H^+}{H}\right)^{-1}
  \left(\frac{N^{\rule{0ex}{1ex}}}{H}\right)
  \left(\frac{N^+}{N}\right)\;T_4^{-0.474}\;e^{-2.18/T_4}\;EM
\end{equation}
where $T_4$ is in units of $10^4$ K and $EM = \int n_e^2\,dl$ is the
emission measure in units of cm$^{-6}$ pc. Note also that $I$ is now
in R. The [\sii]/[\nii] intensity ratio is then
\begin{equation}
  \label{eq:siinii}
  \frac{I_{6716}}{I_{6583}} = 4.69\;
  \frac{\left(\frac{S^{\rule{0ex}{1ex}}}{H}\right)}
  {\left(\frac{N^{\rule{0ex}{1ex}}}{H}\right)}\,
  \frac{\left(\frac{S^+}{S}\right)}{\left(\frac{N^+}{N}\right)}\;
  e^{0.04/T_4}\;T_4^{-0.119}.
\end{equation}
We use the solar (S/H) $= 1.86\times10^{-5}$ from Anders \& Grevesse
(1989\markcite{ag89}) and (N/H) $= 7.5\times10^{-5}$ from Meyer,
Cardelli, \& Sofia (1997\markcite{mcs97}) for the gas-phase abundances
of S and N.

Because of the similar first ionization potentials of N and H (14.5
and 13.6 eV) and a weak charge-exchange reaction, N$^+$/N$^0$ tracks
H$^+$/H$^0$ in photoionization models (Sokolowski 1994\markcite{s94}),
so that in \ha\ emitting regions, where the fractional ionization of H
has been measured to be near unity (Reynolds et al.\ 
1998\markcite{rhth98}), N$^+$/N $\rightarrow$ 1 also. However, with a
second ionization potential of 29.6 eV, N most likely ionizes no
higher than N$^+$. Recent photoionization models run by Howk \& Savage
(1999\markcite{hs99}) over a wide range of temperatures and ionization
parameters show that even if the ionizing spectrum in the WIM is as
hard as $T_e\approx40,000$ K, N$^{++}$/N $< 0.3$.  Furthermore,
Reynolds \& Tufte (1995\markcite{rt95}) and Tufte (1997\markcite{t97})
found that He$^+$/He ranges from 0.3 to about 0.6 in the WIM,
suggesting a spectrum that is significantly softer than 40,000 K, and
therefore implying a very small N$^{++}$/N ratio. Sulfur, on the other
hand, has a low enough first ionization potential, 10.4 eV, that it is
most likely fully ionized in the gas we are studying. In addition, its
second potential, 23.4 eV, lies slightly below the neutral He edge at
24.6 eV. The existence of He$^+$ in the WIM (Reynolds \& Tufte
1995\markcite{rt95}; Tufte 1997\markcite{t97}) and the model
calculations of Howk \& Savage (1999\markcite{hs99}) both imply that
indeed some S will be ionized to S$^{++}$.

In the limit where S$^+$/S and N$^+$/N are both unity and at a typical
WIM temperature of $T_4=0.8$, Equation~\ref{eq:siinii} gives
$I_{6716}/I_{6583} = 1.26$. This relationship is plotted as the solid
line in Figure~\ref{fig:loc-grid}. Dotted lines of lower slope show
the effect of decreasing S$^+$/S (presumably due to an increase in
S$^{++}$) while keeping N$^+$/N $= 1$. In this interpretation, the
data imply that S$^+$/S ranges between about 0.25 and 0.8, with an
average value for the Local diffuse background of 0.6--0.65.  Note
that the location of the \hii\ region symbols (filled triangles and
squares) in Figure~\ref{fig:loc-grid} indicate that these regions have
systematically lower S$^+$/S than the diffuse background. In these
cases the lower ratios are more likely due to the expected behavior
that closer to the ionizing source, where the ionization parameter is
large, more S$^+$ is ionized to S$^{++}$, producing a corresponding
decrease in the observed [\sii] intensity relative to \ha\ 
(Domg\"orgen \& Mathis 1994\markcite{dm94}; Sokolowski
1994\markcite{s94}).

Just as the [\nii]/\ha\ and [\sii]/\ha\ ratios can together give us
information on the ionization state of S, examining the ratios alone
can provide some estimate of the temperature. From
Equations~\ref{eq:itoem} and \ref{eq:nii} the [\nii]/\ha\ ratio is
given by
\begin{equation}
  \label{eq:niiha}
  \frac{I_{6583}}{\iha} = 1.63\times10^5\;\left(\frac{H^+}{H}\right)^{-1}
  \left(\frac{N^{\rule{0ex}{1ex}}}{H}\right)
  \left(\frac{N^+}{N}\right)\;T_4^{0.426}\;e^{-2.18/T_4}.
\end{equation}
Since N$^+$/N $\approx$ H$^+$/H (see above), the H and N ionization
factors can be removed from Equation~\ref{eq:niiha}, with the result
that for a a given (N/H) abundance, [\nii]/\ha\ is only a function of
the electron temperature. As an example, the typical WIM temperature
of $T_4 = 0.8$ then leads to [\nii]/\ha\ $= 0.72$. Lines of constant
temperature are plotted on Figure~\ref{fig:loc-grid} as dashed
vertical lines. With a better estimate for temperature from this line
ratio, we also recompute Equation~\ref{eq:siinii}, including the
temperature dependencies this time, to form the dash-dotted line in
Figure~\ref{fig:loc-grid}. This correction is small.

Figure~\ref{fig:per-grid} shows the same grid of S$^+$/S and $T$
overplotted on the Perseus arm ratios. The limits placed on S$^{++}$/S
in the WIM by Howk \& Savage (1999\markcite{hs99}) are at odds with
the significant number of points below the S$^+$/S $= 0.5$ line in the
Perseus arm. As noted in \S\ref{sec:obs}, there may be an atmospheric
line within the integration range of Perseus arm [\nii] emission. By
comparing [\sii] and [\nii] spectra toward the same locations, we
derive an upper limit of $0.1$ R for the intensity of this line. The
maximum effect of removing this line is displayed for a few sample
points in Figure~\ref{fig:per-grid} by translating the solid sample
error bars points to the location of the dotted error bars. Such a
correction would imply a slightly higher S$^+$/S ratio on average for
the Perseus arm, but not enough to explain the entire difference
between the arms.  We may be probing a harder radiation field in the
Perseus arm gas than their model input spectra or our assumed value of
S/H may be too high.  A reduction in the gas phase sulfur abundance by
0.5--0.75 (0.3--0.1 dex) would bring our inferred S$^{++}$/S fractions
more in line with their models.

The Perseus arm gas appears to have systematically lower S$^+$/S than
the Local arm, with an average value closer to 0.5.
Equation~\ref{eq:siinii} shows that differences in abundances between
the Local and Perseus arm could change the slope of the [\sii]/\ha\ 
and [\nii]/\ha\ relationship.  However, Afflerbach, Churchwell, \&
Werner (1997\markcite{acw97}) measured the Galactic abundance
gradients in ultra-compact \hii\ regions for O, S, and N.  Using their
best fits, N/H falls off faster than S/H with increasing
Galactocentric distance so that S/N is $\sim 5\%$ higher in the
Perseus arm. This is the opposite effect needed to explain the
difference in the Perseus and Local arm emission ratios. More likely,
Figure~\ref{fig:per-grid} suggests that this region of the Perseus arm
has a higher fraction of S$^{++}$ than the Local arm region. Rand
(1998\markcite{rand98}) observed an increase in [\oiii]/\hb\ with
height above the plane of NGC 891 and concluded that an additional
source of ionization is needed beyond photoionization from hot stars
in the disk. In a region where O$^{++}$ ions are being maintained,
S$^+$ becomes a less dominant species. This is consistent with the
inferred differences in S$^+$/S between the Perseus arm
(Figure~\ref{fig:per-grid}) and Local arm (Figure~\ref{fig:loc-grid})
data, since lines of sight pass through the Perseus arm at much higher
$|z|$ than for Local gas. One problem with this interpretation is that
we do not see a substantial fall in the [\sii]/[\nii] ratio with
increasing $|z|$ in the Perseus arm (Figure~\ref{fig:lrvb}). Instead
of an additional ionizing source, it is also possible that there are
marked differences in the strength of the halo ionizing field between
the Local and Perseus arms. For example, NGC 891 shows considerable
variation in the extent of extended \ha\ emission along the galactic
plane (Rand, Kulkarni, \& Hester 1990\markcite{rkh90}; Rand
1998\markcite{rand98}).

The scatter of data points suggests significant variations in the
electron temperature within the mapped region, from about 6,000 K to
10,000 K. The most striking result from these temperature estimates
comes when re-examining the trend of the [\nii]/\ha\ ratio in
Figures~\ref{fig:per-lr}b and \ref{fig:lrvb}. From the rise in this
ratio with increasing $|z|$ and Equation~\ref{eq:niiha} (or the grid
of Figure~\ref{fig:per-grid}), we infer that the temperature in the
halo rises from about 7,000 K at $|z| = 0.75$ kpc ($b \sim
-17\arcdeg$) to over 10,000 K at $|z| = 1.75$ kpc ($b \sim
-35\arcdeg$). If the electron temperature is truly rising with $|z|$
as our results suggest, the calculation of the scale height of \ha\ 
emitting gas (\S\ref{sec:hvsb}) becomes more complicated. If we
confirm this effect with future observations (see
\S\ref{sec:summary}), the net effect (Equation~\ref{eq:itoem}) is that
the true scale height of the gas is larger than the value calculated
in \S\ref{sec:hvsb} by about 20\%.

Note also that since fainter regions tend to have higher [\nii]/\ha\ 
(Figure~\ref{fig:niivsha}), our analysis implies that fainter regions
are hotter. Since $EM \propto \iha T^{0.9}$, higher temperature
regions naturally produce lower \ha\ intensities for a given $EM$;
however, even after converting \iha\ to $EM$ using a varying
temperature estimate from the [\nii]/\ha\ observations, regions with
lower $EM$ are still hotter. Since $EM$ depends on $\phi n_e^2$ and
the pathlength through the emitting region, one of the two must be
smaller in these fainter, hotter regions. Pathlength is an inherently
unphysical parameter to link to temperature and we can think of no
reason why our data set would bias shorter pathlength sightlines
toward higher temperature regions. Thus, we conclude that the hotter
regions most likely have smaller $\langle n_e^2 \rangle$. Since $n_e$
in our mapped region of the Perseus arm decreases with increasing
$|z|$ (see \S\ref{sec:hvsb}), we are not able to ascertain from these
data whether $n_e$ or $|z|$ is the fundamental parameter influencing
the electron temperature.

One possible explanation for the discrepancy between these results and
photoionization models (\emph{e.g.}\ Domg\"orgen \& Mathis
1994\markcite{dm94}; Sokolowski 1994\markcite{s94}) is that at high
$|z|$ or low $n_e$ an additional heating source may dominate over the
heating from photoionization (Bland-Hawthorn et al.\ 
1997\markcite{bfq97}). Since the heating per volume from
photoionization is proportional to $n^2_e$ (it is limited by
recombination), a heating term that is proportional to $n_e$ or did
not depend on $n_e$ at all would become dominant at sufficiently low
densities. Such an effect would decouple the heating from the
ionization of atoms, keeping [\sii]/[\nii] constant, while driving up
the [\sii] and [\nii] intensities relative to \ha. We also note that
if such a process exists in the halo, the increase in [\oiii]/\hb\ 
seen by Rand (1998\markcite{rand98}) in the halo of NGC 891 could be a
direct consequence of this additional heating source, and would not
require a secondary ionization source. Possible heating sources
include, for example, the dissipation of interstellar turbulence
(Minter \& Spangler 1992\markcite{ms97}) and the photoionization of
interstellar grains (Reynolds \& Cox 1992\markcite{rc92}). The idea of
an additional heating source will be explored further in a related
work (Reynolds, Haffner, \& Tufte 1999\markcite{rht99}).

\section{Summary}
\label{sec:summary}

We have presented the first comprehensive, velocity-resolved maps of
[\nii] and [\sii] in the Galaxy. Combining the \ha\ maps from the WHAM
\ha\ Sky Survey in this region, we draw the following conclusions:

\begin{enumerate}
\item We confirm earlier estimates of the distribution of \ha\ 
  emission in the Perseus arm (Reynolds 1997\markcite{rjr97}). With
  continuous coverage from $b = -6\arcdeg$ to $-35\arcdeg$ ($|z| =
  300$ to $1750$ pc at $d = 2.5$ pc), we find that the electron
  distribution derived from the emission is described well by an
  exponential with a scale height of $H = 1.0\pm0.1$ kpc.
\item \mbox{[\sii]}/\ha\ and [\nii]/\ha\ emission in our Galaxy shows
  the same rise with increasing $|z|$ as found in observations of
  edge-on galaxies (Rand 1998\markcite{rand98}; Golla et al.\ 
  1996\markcite{gdd96}).  These ratios appear to be correlated with
  each other over a wide range of emission measures and $|z|$.
\item We suggest that the correlation between [\nii]/\ha\ and
  [\sii]/\ha\ can be attributed to the fact that changes in these
  ratios are primarily due to changes in electron temperature rather
  than ionization parameter, and that scatter in the relationship is
  caused by variations in S$^+$/S. The maps of [\nii]/\ha\ and
  [\sii]/[\nii] (Figures~\ref{fig:loc-lr} and \ref{fig:per-lr}) can
  then be viewed as distributions of electron temperature and
  ionization state, respectively. With this interpretation, we derive
  temperatures in the WIM of 6,000 K to 10,000 K and S$^+$/S ratios of
  0.3 to 0.8. The average value of S$^+$/S appears to be lower in the
  Perseus arm. Higher ionization and lower temperatures are mostly
  found in the two \hii\ regions surrounding the O star $\xi$ Per and
  the B star system $\phi$ Per. Differences in temperature between
  regions may be able to be directly confirmed with observations of
  the faint, higher-level [\nii] $\lambda$5755 transition, which when
  combined with our $\lambda$6483 measurements can directly measure
  the electron temperatures in the emitting gas.
\end{enumerate}

We thank Kurt Jaehnig and Jeff Percival of the University of 
Wisconsin's \emph{Space Astronomy Lab} for their dedicated engineering 
support of WHAM; Nikki Hausen, Mark Quigley, and Brian Babler for 
their contributions to the data analysis; and Trudy Tilleman for 
essential night-sky condition reports from Kitt Peak, which have made 
remote observing possible.  We acknowledge the use of the SIMBAD 
database, operated at CDS, Strasbourg, France.  This work is supported 
by the National Science Foundation through grant AST9619424.

\newpage

\figcaption[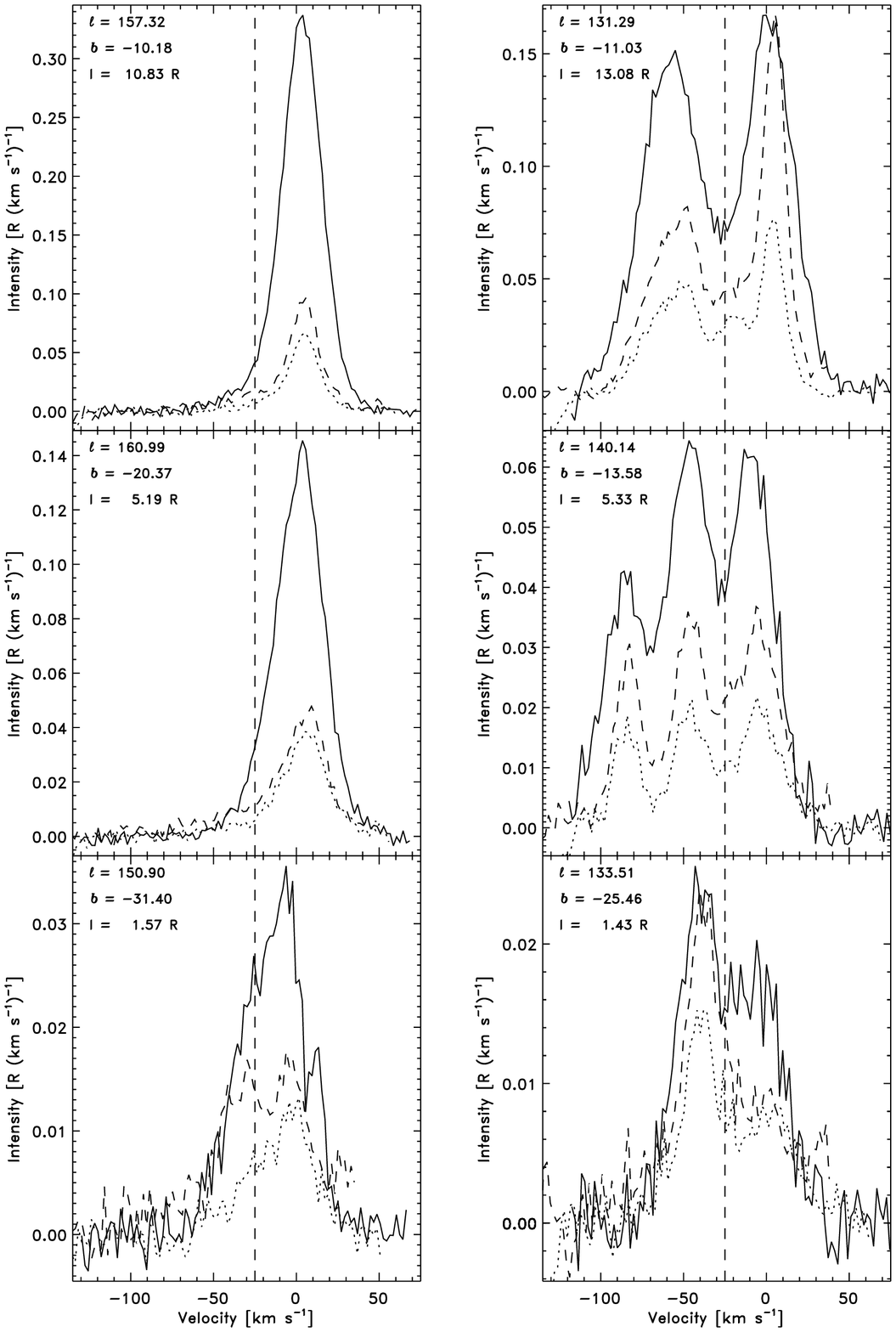]{Sample \ha\ (solid line), [\nii] (dashed
  line), and [\sii] (dotted line) spectra from several directions are
  plotted as intensity versus LSR velocity. The total integrated
  intensity of the emission is labeled in the upper-left corner of
  each plot. The dashed vertical line denotes the delineation for
  integration ranges between the Local and Perseus arms in this paper.
  \label{fig:spectra}}

\figcaption[fig2.ps]{(\emph{a}) \ha\ emission from the local Orion
  Arm.  This pseudo-color image shows the \ha\ emission integrated
  over $\vlsr = -10$ to $+10$ \kms. The axes are Galactic longitude
  and latitude. The overplotted box shows the location of the original
  \ha\ background survey in this region (see text). Emission line
  ratio maps of (\emph{b}) [\nii]/\ha, (\emph{c}) [\sii]/\ha, and
  (\emph{d}) [\sii]/[\nii] are also shown for the same velocity band.
  The color scale in each image is histogram equalized to enhance
  spatial detail. (\emph{Preprint Note:} A color PostScript version of
  this figure can be found at
  \texttt{http://www.astro.wisc.edu/wham/papers.html})
  \label{fig:loc-lr}}

\figcaption[fig3.ps]{(\emph{a}) \ha\ emission from the Perseus Arm,
  approximately 2.5 kpc distant. This pseudo-color image shows the
  \ha\ emission integrated over $\vlsr = -50$ to $-30$ \kms. The axes
  are Galactic longitude and latitude.  The overplotted box shows the
  location of the original \ha\ background survey in this region (see
  text). Emission line ratio maps of (\emph{b}) [\nii]/\ha, (\emph{c})
  [\sii]/\ha, and (\emph{d}) [\sii]/[\nii] are also shown for the same
  velocity band. The color scale in each image is histogram equalized
  to enhance spatial detail. (\emph{Preprint Note:} A color PostScript
  version of this figure can be found at
  \texttt{http://www.astro.wisc.edu/wham/papers.html})
  \label{fig:per-lr}}

\figcaption[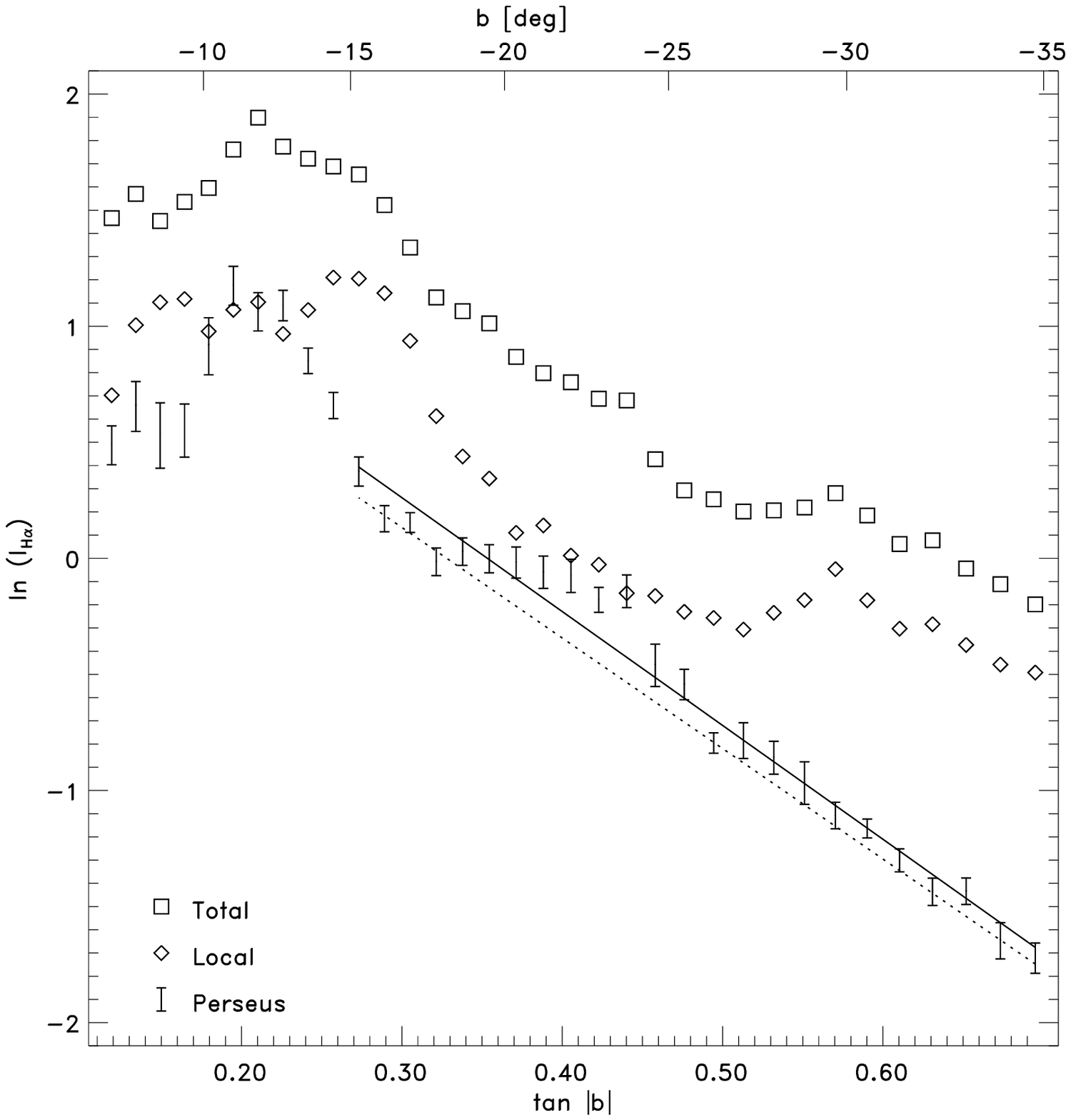]{ The natural logarithm of the median \ha\ 
  intensity between $\ell = 125\arcdeg$ and $152\arcdeg$ is plotted
  versus the tangent of Galactic latitude. Total ($\vlsr = -100$ to
  $+100$ \kms), Local ($\vlsr = -25$ to $+100$ \kms), and Perseus arm
  ($\vlsr = -100$ to $-25$ \kms) emission are denoted by squares,
  diamonds, and vertical bars, respectively. A linear best fit to the
  Perseus arm emission above $b = -15\arcdeg$ is displayed as a solid
  line. The dotted line is a fit to the lower envelope of the data
  (see \S\ref{sec:hvsb}).
  \label{fig:havstanb}}

\figcaption[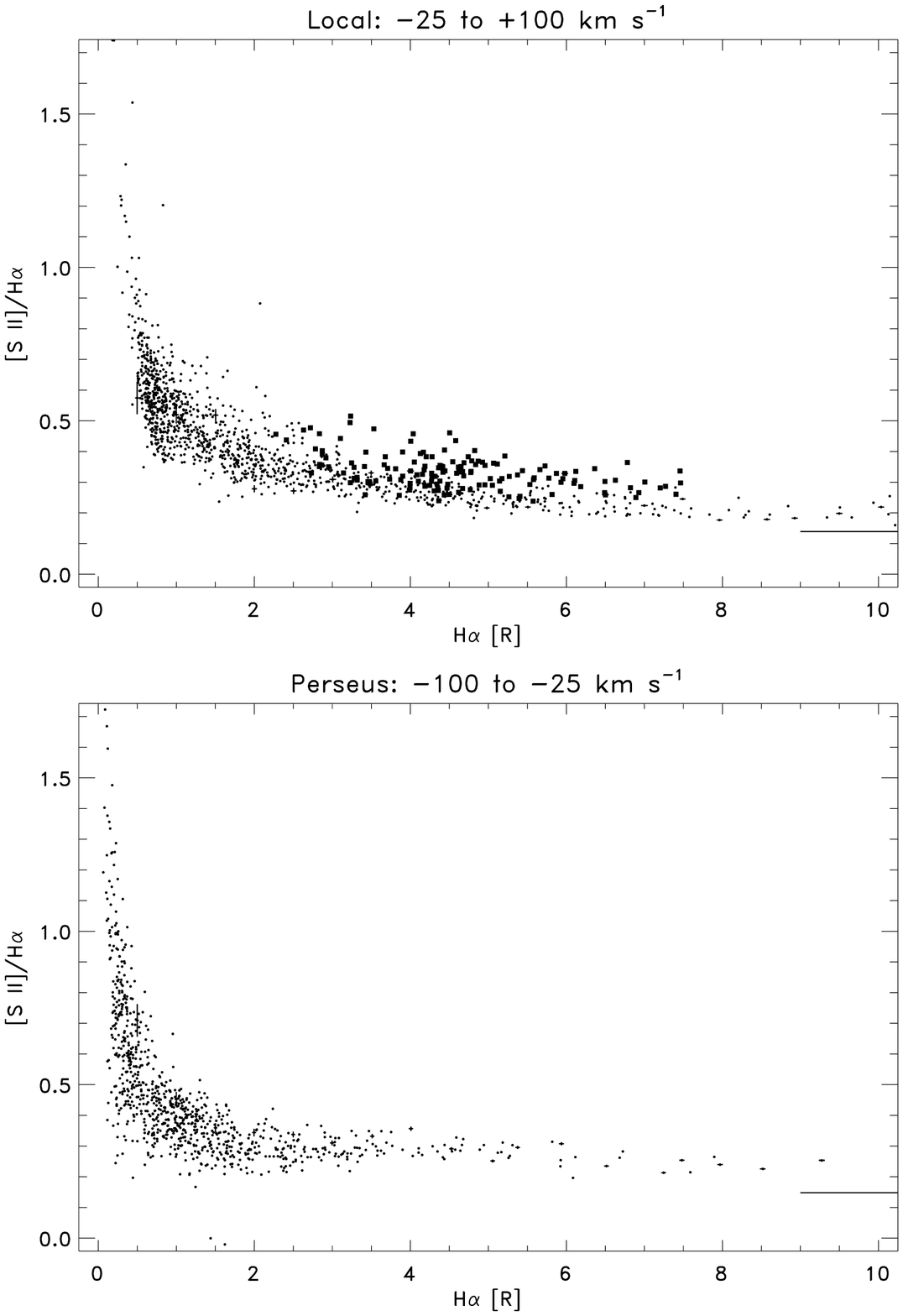]{ [\sii]/\ha\ is plotted versus \ha\ intensity 
  for both the Local and Perseus arm emission.  For the Perseus arm, 
  the longitude range has be restricted to $\ell=125\arcdeg$ through 
  $152\arcdeg$.  Crosses show representative error bars for select 
  points.  The uncertainty in the ratio at 0.5 R is 7-8\%.  Filled 
  squares denote pointings within $6.5\arcdeg$ of the center of the 
  $\phi$ Per \hii\ region mentioned in \S\ref{sec:overview}: 
  $\ell=131\arcdeg$, $b=-11\arcdeg$.  The solid line denotes the 
  average ratio of pointings with $\iha > 50$ R (i.e. ``classical'' 
  \hii\ regions).
  \label{fig:siivsha}}

\figcaption[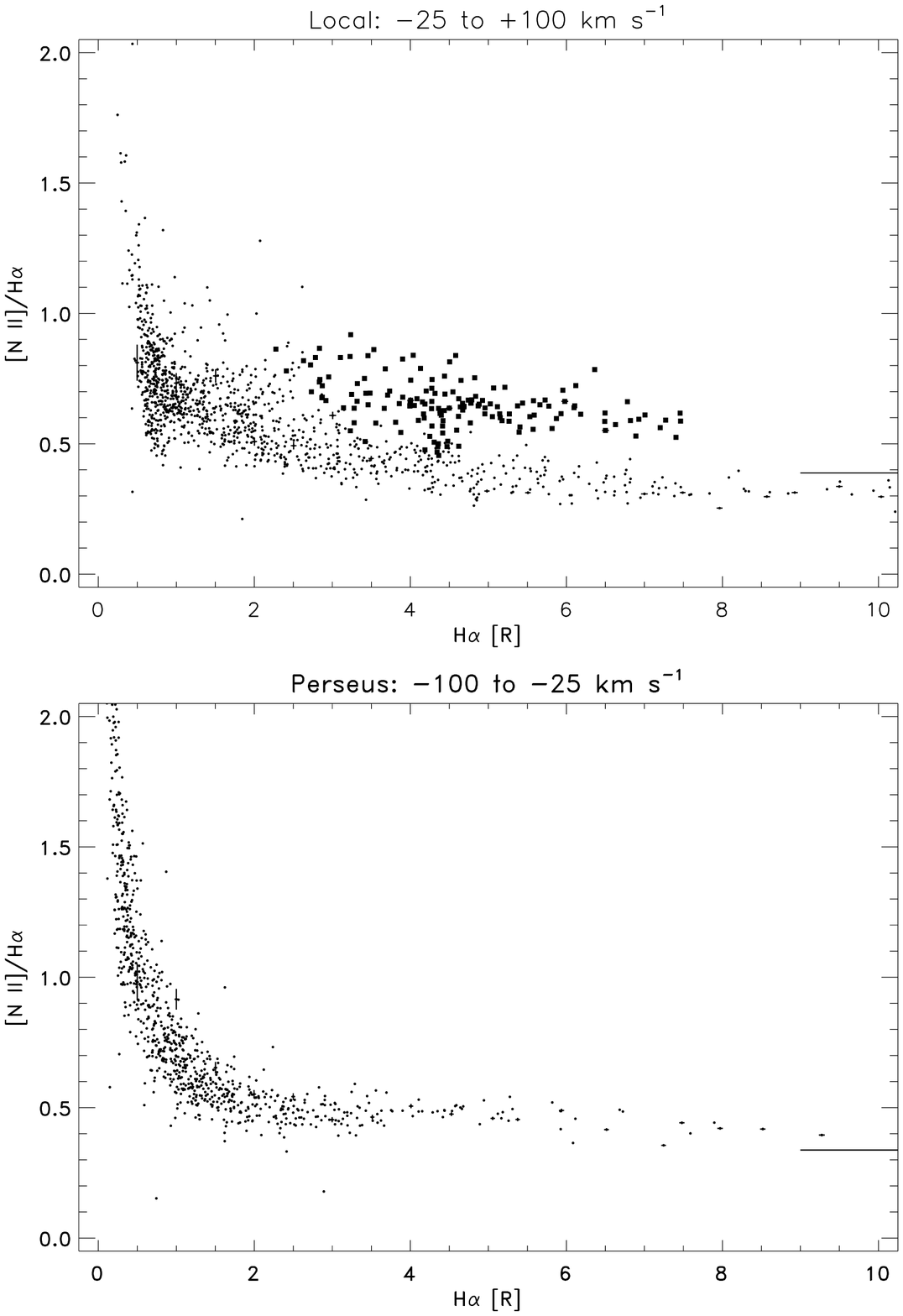]{ Same as Figure~\ref{fig:siivsha}, but for 
  [\nii]/\ha\ versus \ha.
  \label{fig:niivsha}}

\figcaption[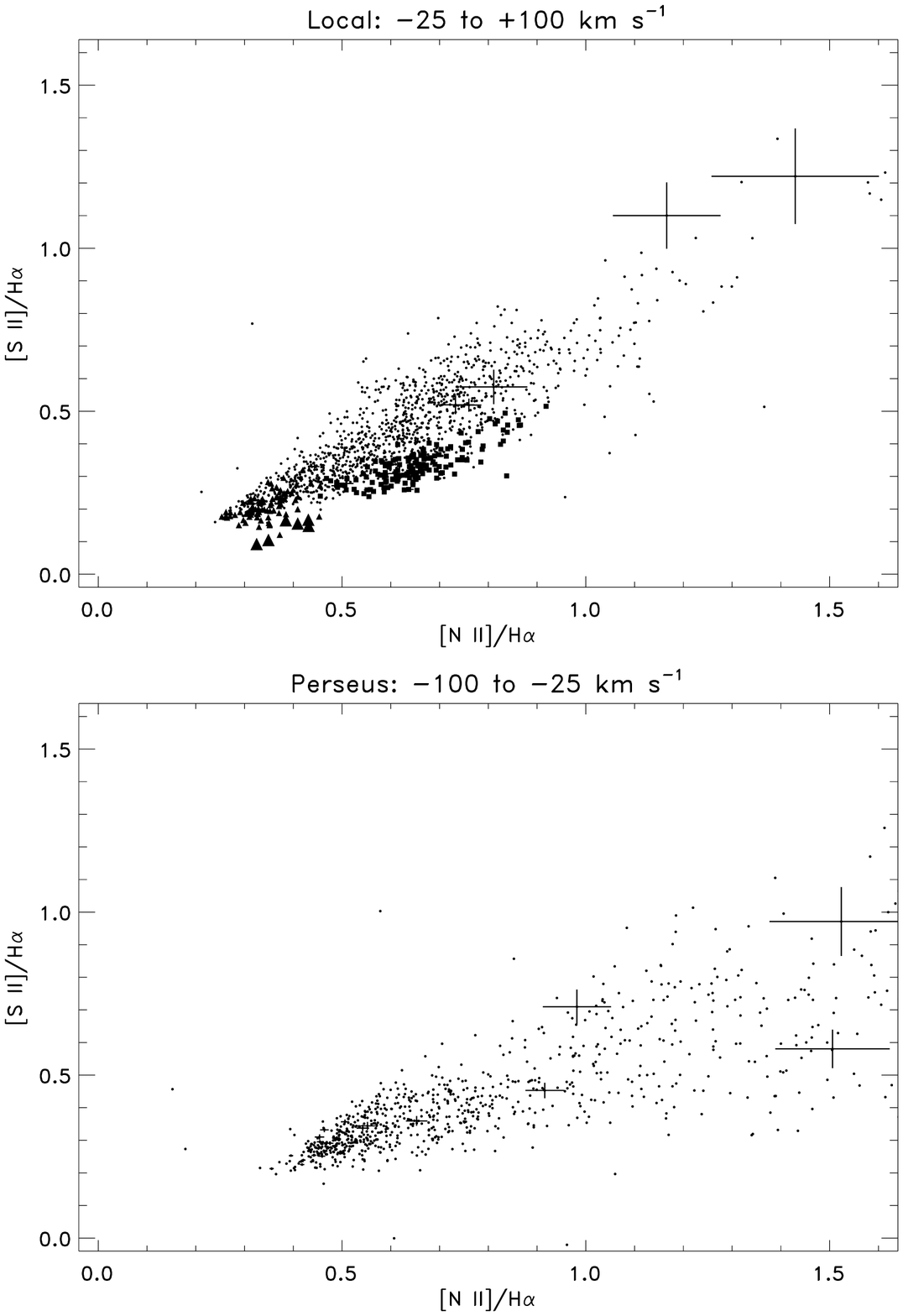]{ Same as Figure~\ref{fig:niivsha}, but for
  [\sii]/\ha\ versus [\nii]/\ha. Additional plot symbols (triangles)
  denote pointings within $5\arcdeg$ of the (rough) center of the
  $\xi$ Per \hii\ region: $\ell=160\arcdeg$, $b=-15\arcdeg$.
  \label{fig:siivsnii}}

\figcaption[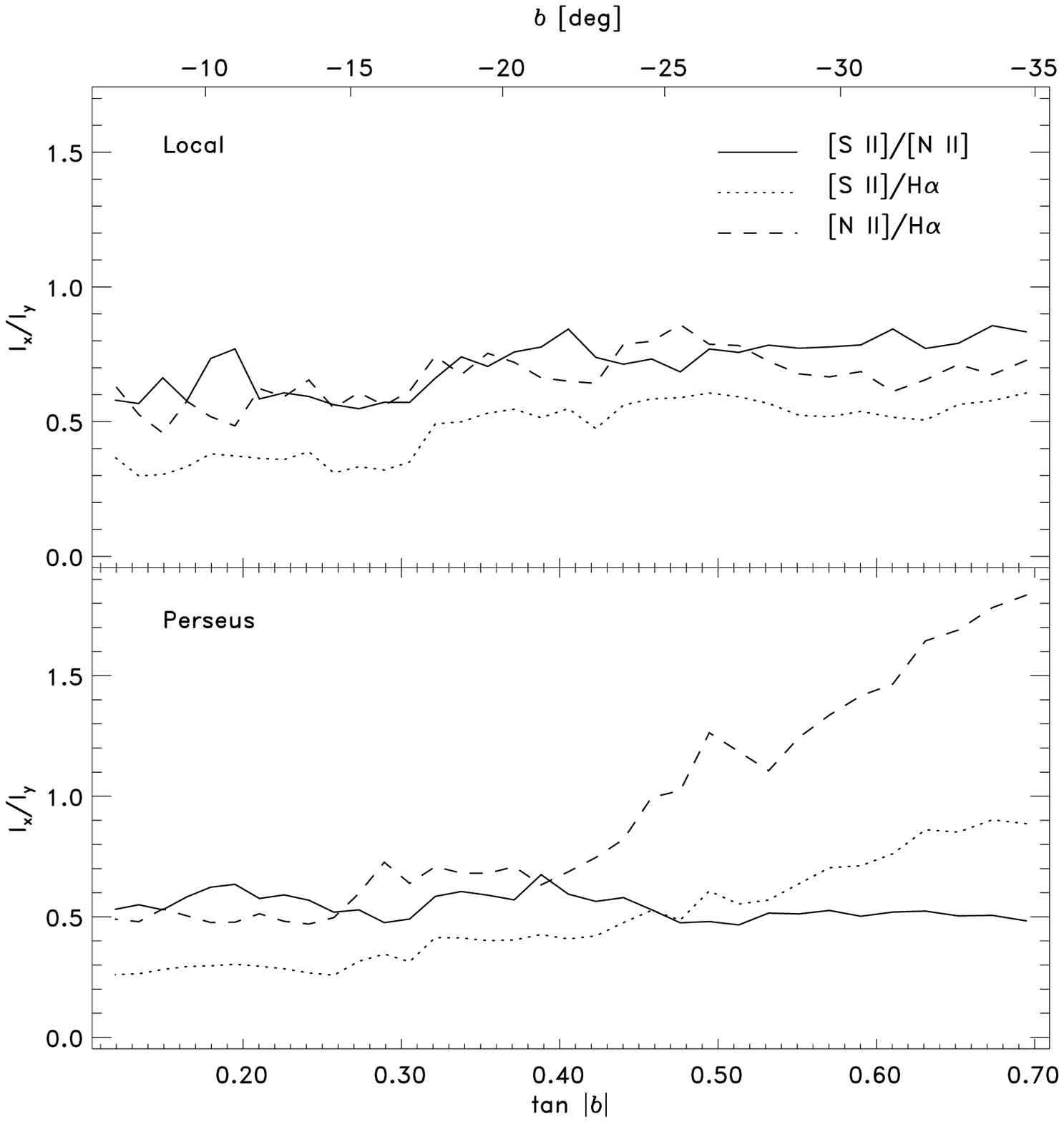]{ Three emission line ratios versus the tangent of
  Galactic latitude: [\sii]/\ha\ (dotted line), [\nii]/\ha\ (dashed
  line), and [\sii]/[\nii] (solid line).  Each data point represents the
  ratio of the median emission between $\ell = 125\arcdeg$ and
  $152\arcdeg$. Intensities have been integrated over the velocity
  range $\vlsr = -25$ to $+100$ \kms\ for the top panel and $\vlsr =
  -100$ to $-25$ \kms\ in the bottom panel.
  \label{fig:lrvb}}

\figcaption[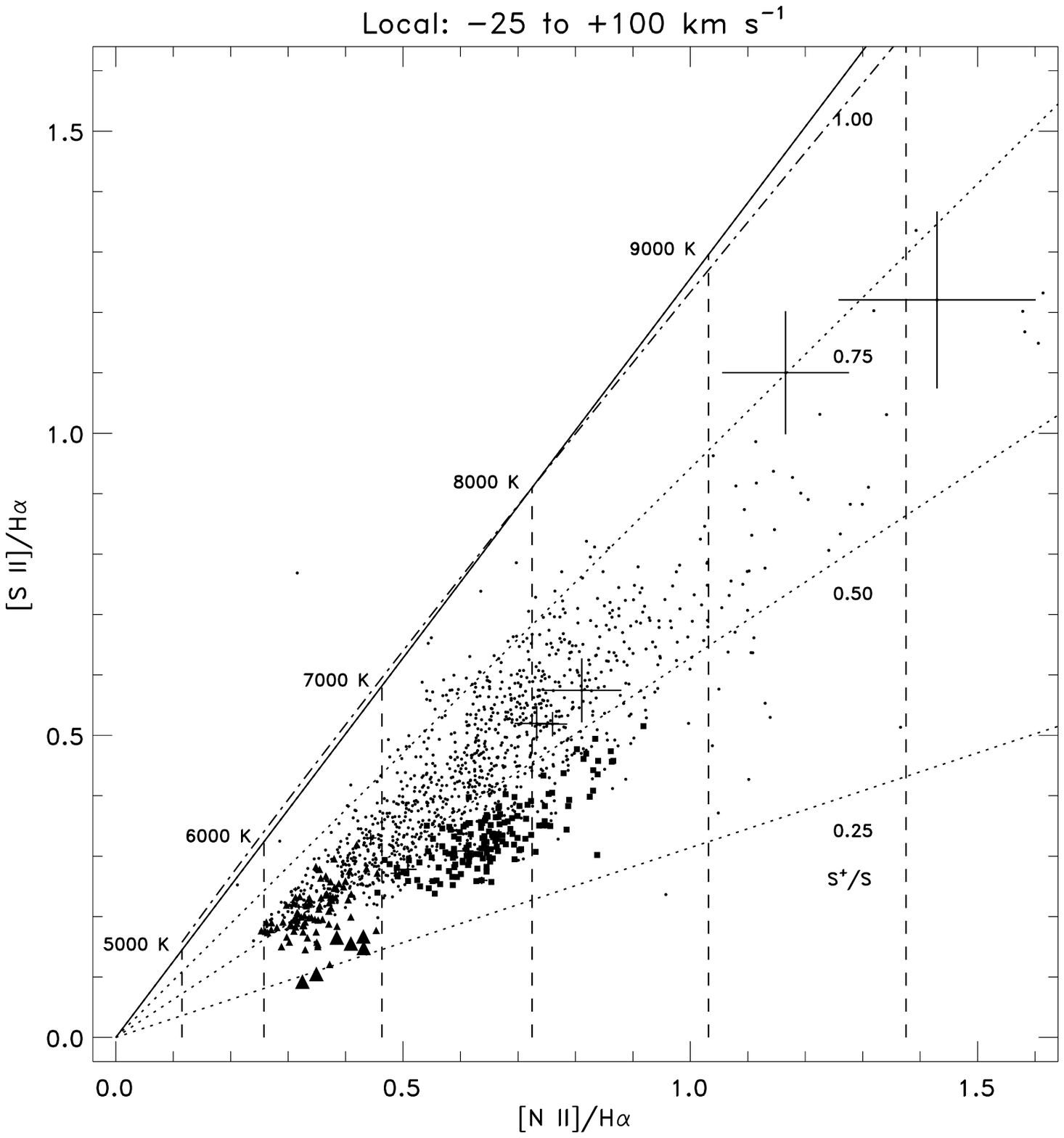]{Same as Figure~\ref{fig:siivsnii}
  for the Local arm only, except with a grid for inferred temperatures
  and S$^+$/S ratios.
  \label{fig:loc-grid}}

\figcaption[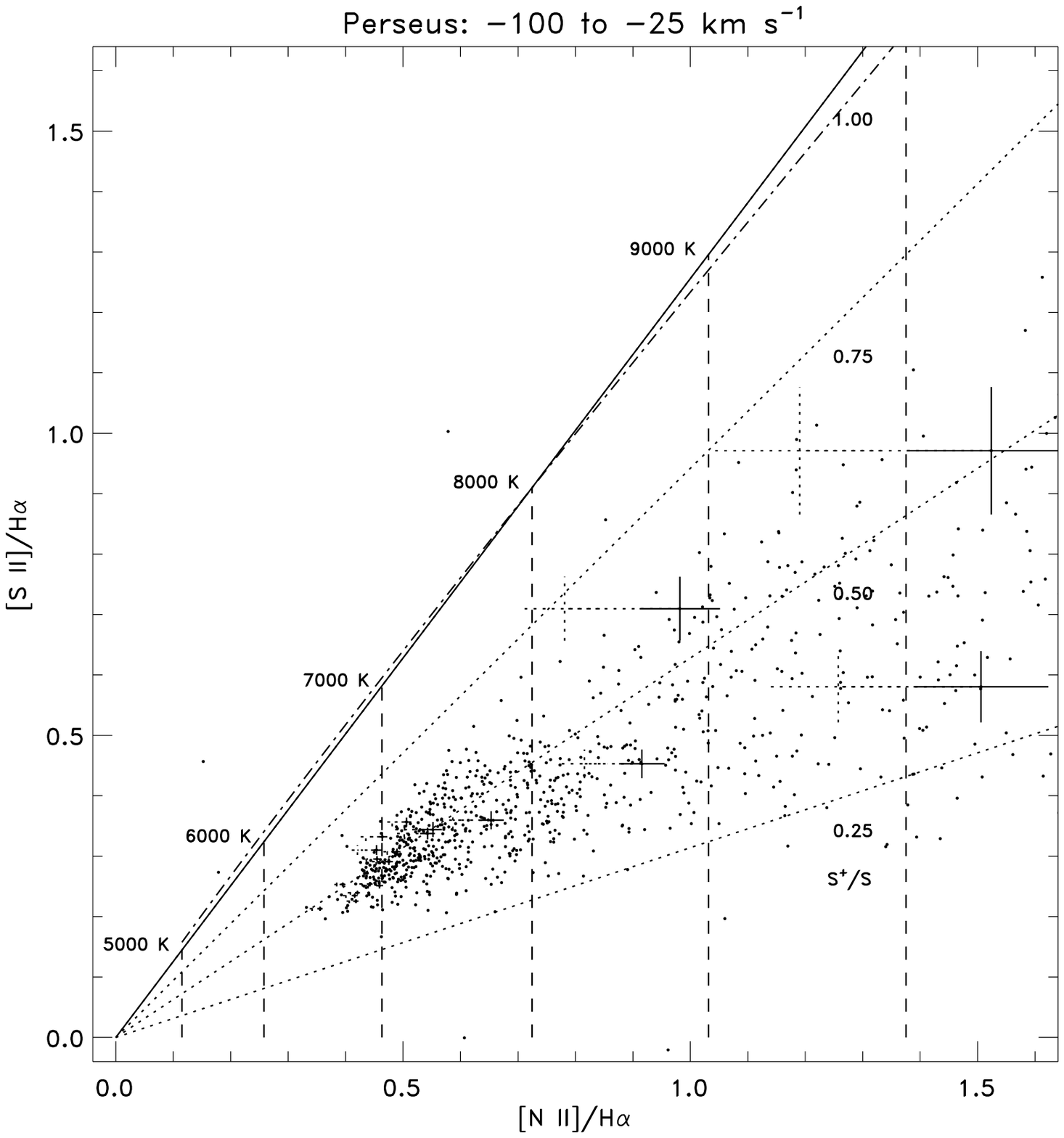]{ Same as Figure~\ref{fig:siivsnii}
  for the Perseus arm only, except with a grid for inferred
  temperatures and S$^+$/S ratios. As before, solid crosses denote
  error bars for representative samples; dotted crosses show how the
  positioning of these data points would change if a weak terrestrial
  line at the velocity of the Perseus arm is as bright as 0.1 R (see
  text).
  \label{fig:per-grid}}

\newpage

\begin{deluxetable}{cccccc}
  \tablecaption{Atmospheric Lines in [\sii] and [\nii]
    Observations\label{tab:atlines}}
  \tablehead{
    \multicolumn{3}{c}{[\sii]} &
    \multicolumn{3}{c}{[\nii]} \nl
    \cline{1-3} \cline{4-6} \nl
    \colhead{Mean\tablenotemark{a}} &
    \colhead{Width} &
    \colhead{Intensity\tablenotemark{b}} &
    \colhead{Mean} &
    \colhead{Width} &
    \colhead{Intensity} \nl
    \colhead{(\kms)} &
    \colhead{(\kms)} &
    \colhead{(R)} &
    \colhead{(\kms)} &
    \colhead{(\kms)} &
    \colhead{(R)} \nl
    }
  
  \startdata
  
  $+26.8$ & $0$ & $0.06$--$0.38$ & $+25$ & $0$ & $0.05$--$0.10$ \nl
  $-72.2$ & $10$ & $0.12$--$0.29$ & $-76$ & $10$ & $0.00$--$0.05$ \nl
  $-113.2$ & $0$ & $0.03$--$0.14$ & $(-40$ & $0$ & $0.05$--$0.10)$\tablenotemark{c} \nl
  
  \enddata
  \tablenotetext{a}{With respect to the geocentric center of the
  emission line.}
  \tablenotetext{b}{Full variation in intensity seen throughout the 
  observation period. Hourly variations are $\sim$ 0.02 R.}
  \tablenotetext{c}{Contaminated by Perseus arm emission. See text.}
\end{deluxetable}

\newpage

\begin{figure}
  \begin{center}
    \leavevmode
    \plotone{fig1.ps}
  \end{center}
\end{figure}

\begin{figure}
  \begin{center}
    \vspace{3in}
    Figure 2 is available separately as a GIF from LANL Preprints or
    as a color PostScript file from
    http://www.astro.wisc.edu/wham/papers.html
    \vspace{3in}
  \end{center}
\end{figure}

\begin{figure}
  \begin{center}
    \vspace{3in}
    Figure 3 is available separately as a GIF from LANL
    Preprints or as a color PostScript file from
    http://www.astro.wisc.edu/wham/papers.html
    \vspace{3in}
  \end{center}
\end{figure}

\begin{figure}
  \begin{center}
    \leavevmode
    \plotone{fig4.ps}
  \end{center}
\end{figure}

\begin{figure}
  \begin{center}
    \leavevmode
    \plotone{fig5.ps}
  \end{center}
\end{figure}

\begin{figure}
  \begin{center}
    \leavevmode
    \plotone{fig6.ps}
  \end{center}
\end{figure}

\begin{figure}
  \begin{center}
    \leavevmode
    \plotone{fig7.ps}
  \end{center}
\end{figure}

\begin{figure}
  \begin{center}
    \leavevmode
    \plotone{fig8.ps}
  \end{center}
\end{figure}

\begin{figure}
  \begin{center}
    \leavevmode
    \plotone{fig9.ps}
  \end{center}
\end{figure}

\begin{figure}
  \begin{center}
    \leavevmode
    \plotone{fig10.ps}
  \end{center}
\end{figure}

\end{document}